\documentclass{llncs}
\usepackage{hyperref}
\usepackage{multirow}
\usepackage{mathtools}
\usepackage{subfigure}
\usepackage{graphicx}
\usepackage{stfloats}
\usepackage{enumitem}
\usepackage{epstopdf}
\usepackage{subfloat}
\usepackage{fancyvrb}
\usepackage{color}
\usepackage{url}

\mathchardef\UrlBreakPenalty=10

\begin{document}

\clubpenalty=10000 
\widowpenalty = 10000

\title{Reading the Correct History? \\ Modeling Temporal Intention in Resource Sharing}

\author{Hany M. SalahEldeen and Michael L. Nelson}
\institute{Old Dominion University, Department of Computer Science\\
Norfolk VA, 23529, USA\\
{hany,mln}@cs.odu.edu}

\maketitle
\begin{abstract}

The web is trapped in the ``perpetual now'', and when users traverse from page to 
page, they are seeing the state of the web resource (i.e., the page) as it exists 
at the time of the click and not necessarily at the time when the link was made.  
Thus, a temporal discrepancy can arise between the resource at the time the 
page author created a link to it and the time when a reader follows the link. This 
is especially important in the context of social media: the ease of sharing links 
in a tweet or Facebook post allows many people to author web content, but the space 
constraints combined with poor awareness by authors often prevents sufficient 
context from being generated to determine the intent of the post.  If the links 
are clicked as soon as they are shared, the temporal distance between sharing and 
clicking is so small that there is little to no difference in content.  However, not all 
clicks occur immediately, and a delay of days or even hours can result in 
reading something other than what the author intended. We 
introduce the concept of a user's temporal intention upon publishing a 
link in social media. We investigate the features that could be extracted from the 
post, the linked resource, and the patterns of social dissemination to model 
this user intention. Finally, we analyze the historical integrity of the shared 
resources in social media across time. In other words, how much is the knowledge 
of the author's intent beneficial in maintaining the consistency of the story 
being told through social posts and in enriching the archived content coverage and 
depth of vulnerable resources?

\end{abstract}

\section{Introduction}
The web is dynamic, and for most users, only the latest version of any particular resource is readily
available.  Although the web does not provide a direct mechanism for
accessing prior states of a resource, these states can be accessed via
web archives like the Internet Archive or in some cases (e.g., wikis)
the web site software might implement a revision control system. However, 
the archival coverage is uneven and few people are aware of this archival existence. 
Thus, a temporal discrepancy can arise between the resource at the time the page's author created a link to it
and the time when a reader follows that link. In other words, \textit{did the
author intend for you to see the web page as it existed when they
shared it ($t_{tweet}$), or did they intend for you to see the version
as it existed at the time the reader clicked on the link
($t_{click}$)?}

If the period of time between the sharing and the clicking events is small, 
in most cases there will be no tangible difference. However, the more time that elapses between those two events, the greater the possibility of content change jeopardizing the 
consistency between the social post and the shared content.

If social media is supplanting journalism as the ``first rough draft of
history'', then we cannot assume the time between sharing and clicking
will be so small that the gap can be ignored.  In preliminary research
we have discovered after just one year, tweets about the Egyptian
Revolution have lost (i.e., HTTP 404) approximately 11\% of the
resources they link to \cite{ws-dl:blog:losing}.  Furthermore, many of those that remained
(i.e., HTTP 200) were no longer what the original author intended. If we consider such posts as part of the
historical record (i.e., a library), then the pages that referenced in these posts are
part of the historical record as well.  If they are not preserved in
the manner in which they were intended to be shared, then we are losing
pieces of history.  Many of the pieces needed to resolve this problem
are in place: there is a growing infrastructure of web archives and
the protocols to access them. Social media such as tweets provides
uneditable creation dates to mark the sharing event, and often provides
personalized, and unique URI aliases for the shared resource.  All of
these can be combined to create the proper context for determining
the correct temporal intention.

\section{Problem Definition}\label{definition}
Every day, millions of pictures, videos, links, and tweets are shared between social media users all over the globe. 
Those social posts differ in purpose and expected audience. Some posts are made to convey mood, personal state or activity, 
opinion about a certain topic, humor, express anger, share useful information, or even pranking and spam. The author of a 
post is creating web content which may link to one or more other resources as well. These resources could be a web page, 
media file, another social post, or a document. While time passes, the content which the author 
created remains unchanged while the linked resources do not maintain the same stability 
as in most cases those resources are out of the author's control. In several cases, the shared resources go missing and we analyzed the 
percentages of missing shared resources as a function of time in earlier work \cite{TPDL2012:Losing}. 
However, in several other cases the linked resource is still on the live web but it has changed and no longer relevant to what 
the author intended to convey.

\begin{figure}[ht!]
     \begin{center}
        \subfigure[A tweet depicting Obama's news conference and the topic of the Haitian Earthquake]{%
            \label{fig:haiti1}
            \includegraphics[scale=0.42]{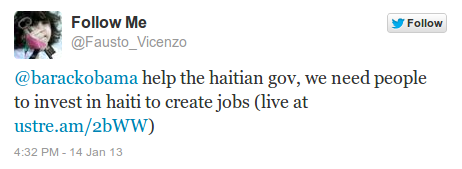}
        }\\
        \subfigure[The state of the embedded resource at the time of clicking depicting the 2013 Superbowl]{%
           \label{fig:haiti2}
           \includegraphics[width=0.32\textwidth]{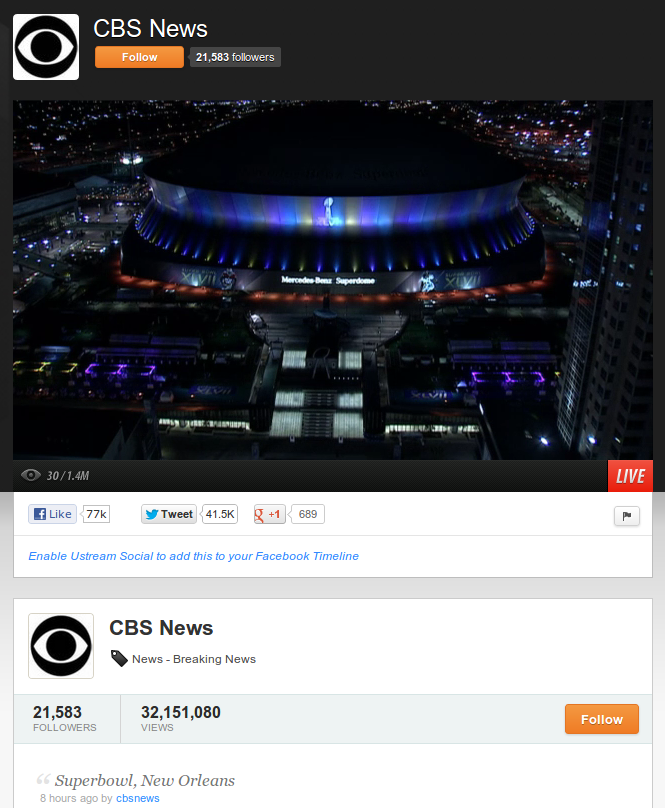}
        }\\
        \subfigure[Using the twitter expanded interface showing a third state of the resource.]{%
            \label{fig:haiti3}
            \includegraphics[width=0.32\textwidth]{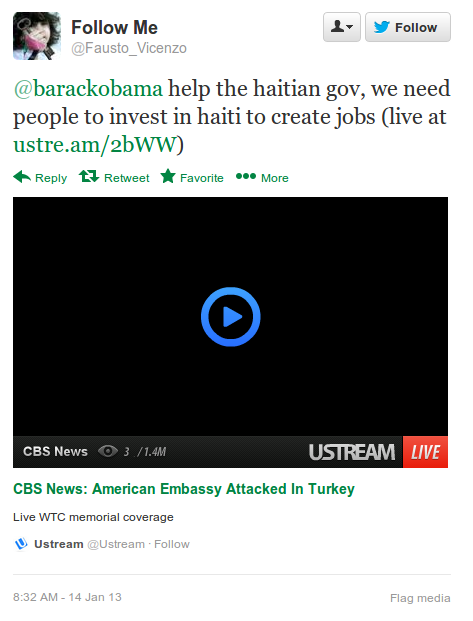}
        }
    \end{center}
\vskip 5mm
    \caption{Different resource states at $t_{tweet}$ and $t_{click}$.}%
   \label{fig:example}
\end{figure}

This change could be tolerated or have reduced effect if it was on an individual level. To elaborate, the effect of the change 
on a tweet depicting a family's cat is magnitudes lower than a change of a tweet showing a police officer pepper-spraying the faces of 
peaceful protesters (e.g., Occupy Wall Street). Losing the consistency of the former tweet affects only the individuals related 
to the family while losing the consistency of the latter more directly affects our cultural historical record.

To further explain the problem, let us examine the following scenario. On January the 14th, President Obama held his final news 
conference of his first term in the East Room of the White House. He discussed several issues among which was 
the third anniversary of the earthquake disaster in Haiti. On the same day, a user tweeted about it while watching the speech as 
shown in figure \ref{fig:haiti1}. Clicking on the link associated with the tweet, a page is rendered depicting a stream from the 
Mercedez-Benz superdome in New Orleans, Louisiana covering the Superbowl American football game of 2013 as shown in figure 
\ref{fig:haiti2}. This indicates a mismatch between the resource state at $t_{tweet}$ and $t_{click}$. The text of the tweet makes it clear that the resource state at $t_{click}$ was not the author's intention. Furthermore, using the new Twitter interface to expand the tweet and see the cached caption and embedded 
linked resource, we witness even a third mismatch. Figure \ref{fig:haiti3} shows the cached caption pointing to a story about the attack 
on the American Embassy in Turkey on the 2nd of February 2013.

A possible solution would be to estimate the
author's temporal intention (either the state of the resource at the time
of the tweet $t_{tweet}$ or the state of the resource at the time of reading) upon reading a tweet,  and
recommend to the user either an archived version of the resource at the
closest time to the publishing timestamp, or the current version on the
live web. Furthermore, we could preemptively push a copy of the resource
at the time of the tweet into a web archive so that the intention is fully
preserved.
\section{Related Work}
Intention, mood, and sentiment have been analyzed in different
contexts, but not with respect to time. Furthermore, this research
builds on a large body of work involving detecting changes in web pages,
archiving, and studying social media.

The web is ever-changing and what one might
share or post today might change or disappear tomorrow. Losing web
resources and finding them again has been the scope of several studies.
For digital libraries, Nelson and Allen analyzed the persistence and
availability of objects in a digital library
\cite{DBLP:journals/dlib/NelsonA02}.  From the aspect of web decay
Bar-Yossef et al. \cite{Bar-Yossef:2004:STG:988672.988716} proposed a measure of decay and algorithms to compute it efficiently.
Consequently, Klein and Nelson analyzed the loss and rediscovery of
websites to pinpoint the reasons behind this behavior
\cite{Klein:2008:RLS:1429852.1429903}.

The problem of disappearing or changing resources has been
well-studied.  The changing aboutness of live web pages has been
studied in the Walden's Path project \cite{996387} and the link vetting system \cite{Dai:2009:VLW:1645953.1646220}.  For link
rot, Kahle originally reported the expected lifetime of a web page is
44 days \cite{kahle:preserving}.  Loss of references and URIs appearing
in the academic literature have been studied numerous times, with exact
loss rates varying depending on the corpus \cite{or2011:sanderson}.  In our ``Just-in-time'' preservation research we
discovered new locations of web pages that are missing in the current
web \cite{klein:thesis}.  We investigated a variety of techniques,
including using page titles \cite{ht10:goodtitle}, tags \cite
{DBLP:conf/ercimdl/KleinN11}, and lexical signatures
\cite{Klein:2011:RMW:1998076.1998101}, all of which
could be used as queries to search engines to find replacement copies
of the missing web page.

Computing the change rates of web resources is a well-studied
phenomena.  Cho and garcia-Molina studied the change rate of web pages to determine the
best policies for web crawlers \cite{857170}, as well as
studying how to handle late arrivers in a collection \cite{988674}.
Other studies have been done about understanding the web content dynamics \cite{1498837} and upon which to develop the crawl policies for enhancing archival coverage \cite{AWUPCP:2011:MBS}.  

Due to the tremendous growth of the social media
\cite{facebook,twitter} and the continuous expansion and addition of
new social network-based applications on the
web \cite{citeulike:336118}, a significant body of
research has been created specifically to analyze social media networks from different
angles. For example, the use of URI shorteners, especially
with respect to their use in social media, was studied in
\cite{antoniades2011we}. There has been significant progress recently in sentiment analysis
and gauges for public and individual mood, especially using Twitter feeds
and blog content. Twitter,
specifically, has been analyzed for collective sentiment thoroughly
where mood transition observed in Twitter
\cite{Mogadala:2012:TUB:2390131.2390145} has been utilized in politics
\cite{Bermingham_onusing}, stock market
\cite{DBLP:journals/corr/abs-1010-3003} and others. Intention analysis and detection in web science have several flavors
and can be found in different contexts.  It was analyzed as an
independent concept
\cite{Jethava:2011:SMU:2009916.2009971},
in data mining \cite{Chen:2002:UIM:598693.598776}, in query intent
analysis \cite{Jansen:2007:DUI:1242572.1242739},
in user click models for search \cite{Li:2008:LQI:1390334.1390393},
in search result diversification \cite{Santos:2011:ISR:2009916.2009997}, in cluster
analysis \cite{quteprints47855}, in spam and phishing attacks detection
\cite{Wu:2006:WWP:1143120.1143133}, and in
microblogging \cite{Java:2007:WWT:1348549.1348556}. To our knowledge, 
there is no published research describing temporal
intention in the context of web navigation and social media
dissemination.

In regards to data collection, we are in need of a large data set that
captures human temporal intention. To collect this, prior and during
the phases of experimental design, we examined several publications
depicting crowd sourcing \cite{Tian:2012:LCP:2339530.2339571} and most
specifically Amazon's Mechanical Turk
\cite{Elsas:2010:LTD:1718487.1718489} which has been used in generating
ground truth data for a similar-scoped study in detecting music
moods \cite{Lee:2012:GGT:2232817.2232842}.

As for the archiving aspect of our study, the existence of Memento, TimeMaps, and multi-archive aggregators has
greatly facilitated research with archives. The motivation for the Memento Framework \cite{nelson:memento:tr} is achieving a tighter
integration between the current web and remnants of the web of the
past. Archival versions (or mementos) of web resources do exist, both in
special-purpose web archives such as the Internet Archive and the
on-demand WebCite archive, or in version-aware servers such as content
management systems (CMS, e.g.  Wikipedia) and version control systems.

\begin{table*}
\begin{center}
\begin{tabular}{crc|c}
& & \multicolumn{2}{c}{Tweet and resource are:} \\
& & relevant & not relevant \\
\cline{3-4}
\cline{3-4}
\multirow{2}{*}{Linked resource has:}
& changed & $t_{click}$ & $t_{tweet}$ \\ 
\cline{3-4}
& not changed & $t_{tweet}$ & either or undefined \\ 
\cline{3-4}
\end{tabular}
\vskip 5mm
\caption{\label{tab:model}Temporal Intention Relevancy Model.}
\end{center}
\end{table*}
\section{Crowd Sourcing User Intention}
To have a better understanding of a user's temporal intention, we performed
several experiments on Amazon's Mechanical Turk. Subsequently, we discovered that classifying temporal intention is difficult for Mechanical Turk workers. This, in turn, has influenced us to seek a transformation of the problem to another 
domain while maintaining the  semantic consistency as shown in the next sections.

\subsection{Preliminary Work}
Initially, we attempted using Mechanical Turk directly in classifying intention. Our first set of experiments involved sampling 1000 tweets from the Stanford
Network Analysis Project
(SNAP\footnote{\url{http://snap.stanford.edu/}}) Twitter data set. The first step was to prove that Mechanical Turk could be used in representing manually assigned classes of intention made by experts in the field. The classes targeted were 
as follows: did the author of the tweet intended the ``Current State'' of the resource for the reader at any time or the ``Past State'' of the resource at 
the time of the tweet? Or there not enough information?

To achieve this, from the set of 1000 tweets we constructed the ground truth responses for 100 tweets forming the gold standard dataset. The collection 
of the gold standard dataset was performed by
polling via email the members of our Web Science and Digital Libraries
(WSDL) research group and asking them to classify the intention of the intention of the author of a tweet as
either the current version ($t_{click}$), the archived version (past)
($t_{tweet}$), or unknown by looking at the tweet.  The reliability of agreement
within our group of 12, all of whom are aware of web archiving, was
surprisingly low (Fleiss' $\kappa$ = 0.14).  We ran the experiments in
Mechanical Turk, acquiring five evaluations for each of the same 100 tweets from the gold standard dataset.
Similarly, the inter rater between the Mechanical Turk workers was even lower (Fleiss' $\kappa$ = 0.07).
\begin{equation}
Vote_{MT}(tweet)=
\begin{dcases}
\displaystyle \text{Current},&\text{if } \frac{\Sigma Vote_{current}}{N_{turkers}} > k \\
\displaystyle \text{Past},&\text{otherwise}
\end{dcases}
 \label{eq:mt1}
\end{equation}
The threshold \textit{k} in equation \ref{eq:mt1} defines the vote cut off. In this case, \textit{k} = 0.5 as we applied a simple majority vote in deciding the collective vote of the Mechanical Turk workers
(i.e., whichever classification received three out of five voters), and similarly within the 12
WS-DL members. Treating each group as a single entity, the aggregated votes from each of the two datasets were used to calculate the inter rater 
agreement resulting in Cohen's $\kappa$ = 0.04, indicating slight agreement. This slight agreement was yet not sufficient to proceed with our study.
Examining the selection from the SNAP data set, we decided that 
too many of the tweets had vague contexts and were hard to classify.

Given the unclear contexts that were present in the first sample set,
we then tried a richer set from which to sample.  We used the tweets
from the six historical events described in \cite{TPDL2012:Losing}.  For 100 tweets, we built a web page with an
image snap shot of the current version of the page, and a version of
the page closest to $t_{tweet}$ that could be found in a public web
archive.  We held a face to face meeting with our WSDL research group to
determine the ground truth: for each tweet we went around the table and
argued for whichever version we thought matched the author's temporal
intent.  We knew this data set would be biased toward $t_{tweet}$
because most of the tweets described historic, cultural events from
2009-2011.  After deliberation, we arrived at: 82\% past, 9\% current,
and 9\% undecided as our gold standard for this data set.  When we
submitted the jobs to Mechanical Turk, we defined levels of three,
five, seven, and nine evaluations for each tweet.   In the case where we had 
nine evaluations for each tweet, the Mechanical Turk workers would match
our gold standard 58\% of the time if we allowed 5-4 splits.  If we were 
more discerning and counted agreement only in cases where workers agreed
6-3 or better, then the agreement with Mechanical Turk workers fell to 31\%
(and similarly for rating levels three, five, and seven).  

In short, if we required clear agreement on the part of Mechanical Turk
workers, then we did much worse than simply flipping a coin -- in a data
set with a clear bias toward $t_{tweet}$ because of the focus on past events.
It was at this point we decided our approach in guessing the author's 
temporal intent was simply too complicated for Mechanical Turk workers. 
\begin{figure*}[ht!]
     \begin{center}
        \subfigure[Changed and Relevant]{%
            \label{fig:first}
            \includegraphics[width=0.5\textwidth]{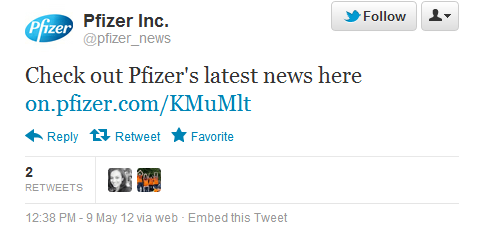}
        }%
        \subfigure[Changed and no longer Relevant]{%
           \label{fig:second}
           \includegraphics[width=0.5\textwidth]{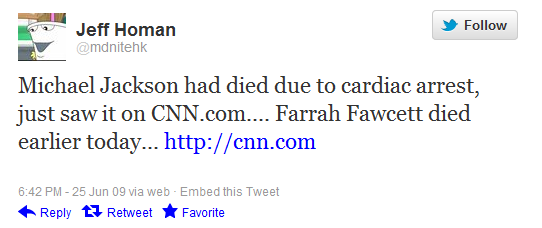}
        }\\ 
        \subfigure[Not changed and Relevant]{%
            \label{fig:third}
            \includegraphics[width=0.5\textwidth]{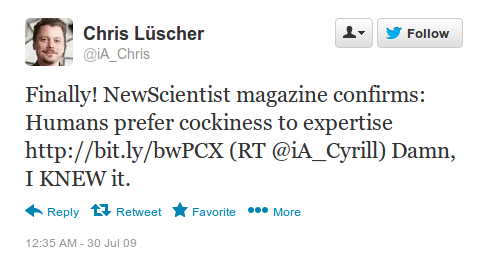}
        }%
        \subfigure[Not Changed and not Relevant]{%
            \label{fig:fourth}
            \includegraphics[width=0.5\textwidth]{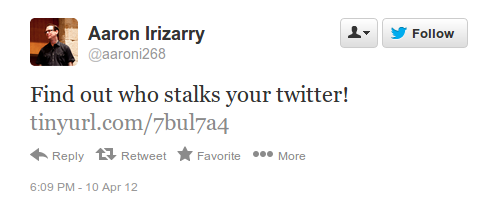}
        }%
    \end{center}
    \caption{Examples of the relevancy mapping of TIRM.}%
   \label{fig:subfigures}
\end{figure*}

\subsection{Temporal Intention Relevancy Model}
To reach our goal of modeling users' temporal intentions, we need to collect 
a large dataset which is not, as discussed in the previous section, a trivial 
task. The difficulty in acquiring the data resides in generating
the ground truth or gold standard for the temporal intention of the
user who authored the original social media post. Initially, our intention was to
generate a small set of gold standard data (e.g., links classified as
representing the user's intention to be either ``the resource at
$t_{tweet}$'' or ``the resource at $t_{click}$'').  We eventually decided that the notion of ``temporal
intention'' was too nuanced to be adequately conveyed in the
instructions for the workers of Mechanical Turk.  Learning from our
previous unsuccessful attempts, we chose to cast the problem
of ``temporal intention'' to one of relevancy between the tweet and the
resource as it exists now.   

Table \ref{tab:model} presents the Temporal Intention Relevancy Model (TIRM) that we will use to inform our
interaction with the workers at Mechanical Turk. To resonate with one of the common types of experiments in 
it, we designed our new experiment as a categorization of relevance problem which the workers are familiar 
with. In each Human Intelligence Task or HIT, the worker is presented with the full tweet, its publishing date,
and in an embedded window, a snapshot of the page that the tweet links to in its current state. Instead of asking workers about
temporal intention of the original author, and possibly confusing it
with the temporal intention of them as a reader, we asked a simpler question ``is this page still relevant to this tweet?''. There is
considerable precedence in the Mechanical Turk community for making
relevance judgements as categorization problems are commonly available as HITs. 

To explain this mapping from intention space to relevancy space, let us assume we have a 
resource \textit{R} which has been tweeted by some author at time $t_{tweet}$. The state of the 
resource at $t_{tweet}$ is $R_{tweet}$. Consequently, another user clicked on the resource to read it at a later time $t_{click}$. The state of the 
resource at $t_{click}$ is $R_{click}$. The rationale for the model is:
\begin{description}
 \item[Changed \& Relevant:] If the resource has changed (i.e.,\\ $R_{tweet}$ is not similar to $R_{click}$) and it is still relevant to the tweet, then there is a strong indication that the
temporal intention of the author must have been the resource as it exists at $t_{click}$ ($R_{click}$). Figure \ref{fig:first} shows an author tweeting about the 
latest updates for a newsletter. The linked resource in the tweet continually changes while the tweet is always relevant to it. This indicates that the author's temporal 
intention is a \textit{current} one.

 \item[Changed \& Non-Relevant:] If the resource has changed and it is not relevant to the tweet, we assume initial relevance and thus the original author 
must have meant to share the resource in the state as it existed at $t_{tweet}$ which is $R_{tweet}$ not $R_{click}$. Figure \ref{fig:second} shows an author 
tweeting about specific breaking news on CNN.com's first page, which by definition changes frequently. This indicates that the author's 
temporal intention to be the \textit{past} version. 

 \item[Not Changed \& Relevant:] If the resource has not changed and it is still relevant to the tweet, 
then we claim that the intention of the author was to share the resource as it existed
at $t_{tweet}$ ($R_{tweet}$), but it is just a fortunate coincidence that the resource 
has not changed and is thus still relevant. Figure \ref{fig:third} shows an author tweeting about an article which still exists. Surely, there is a possibility that the resource 
could change in the future and become non-relevant. This indicates that the author's intention was a \textit{past} one.

 \item[Not Changed \& Non-Relevant:] If the resource has not changed and it is not relevant to the tweet, then we can not be sure of 
the intention and either $t_{click}$ or $t_{tweet}$ will suffice.  This scenario
can occur in spam, mistaken link sharing, or more likely that relevancy relies
on out-of-band communication between the original author and the intended readers\footnote{The Internet meme of ``Rickrolling'' http://en.wikipedia.org/wiki/Rickrolling is a humorous example of purposeful non-relevancy between the context of the link and the link which is to the 1987 pop song by Rick Astley; the point is to ``trick''
users into expecting one thing and the link delivers the song.}.
\end{description}

\subsection{Gold Standard Dataset}\label{gold}

After laying the basis of the intention-relevance mapping in TIRM, we must collect a large body of data to be utilized in the modeling and 
analysis phases. Since we are modeling human intention and mapping it to relevance judging, we will utilize Amazon's Mechanical Turk in collecting the training data. 
However, prior to collecting the training dataset we need to be confident in the ability of our data collection experiment in representing 
the real-life educated judgement. To achieve this goal we created a gold standard dataset by obtaining a small dataset and assigning it to members of our research group, whom we have confidence in their ability to perform the task accurately, and then assign the same dataset to workers in Mechanical Turk. We collect both sets of 
assignments and compare their similarity to ensure the ability of the workers to mimic the judgment of the experts. Mechanical Turk HITs are considerably cheaper, 
easier to manage, and faster to conclude than the expert assignments.

Engineering a relevance HIT for Mechanical Turk's workers was fairly straightforward. For the 
gold standard dataset we randomly picked 100 tweets from the SNAP dataset dating back to June 2009 
and posted them to be classified as ``still relevant'' or ``no longer relevant''. As mentioned earlier, for each HIT we posted the tweet, the date, 
and a snapshot of the resource at $t_{click}$ ($R_{click}$). The experiment requested five unique 
raters with high qualifications (more than 1000 accepted HITs and more than 95\% acceptance rate). Each HIT cost two cents and a maximum time span of 20 minutes. The experiment was completed 
within the first hours from posting and the average completion time per hit was 61 seconds. We examined the data from the workers and dismissed all the 
HITs that took less than 10 seconds indicating a hastly decision. We also filtered out workers who exhibited low quality repetitive assignments and banned them. 
For the same 100 tweets, we invited our research group again to perform this same experiment of relevance. Their assignments have been collected along with the ones from the workers. 
The results are shown in table \ref{tab:turk} showing an almost perfect agreement with Cohen's $\kappa$ = 0.854.

Given this substantial agreement between the gold standard and the workers, we can claim that Mechanical Turk can be used in estimating 
the content's time relevance and in turn to gauge the author's temporal intention after utilizing TIRM. The next step is to expand our dataset and 
collect a larger dataset, for training and testing, to utilize it in the modeling process.

\begin{table}
\begin{center}
\begin{tabular}{|l|l|}
\hline
Agreement in three or more votes & 93\% \\ \hline
Agreement in four or more votes &80\% \\ \hline
Agreement with all five votes & 60\% \\
\hline
\end{tabular}
\vskip 5mm
\caption{\label{tab:turk}Agreement between the research group and Mechanical Turk workers for 100 tweets.}
\end{center}
\end{table}
From the SNAP dataset of tweets we extracted a large number of tweets starting from June of 2009 at random. For a social media post, in this case a tweet, we want to acquire as much data as possible about 
its existence such as content, age, dissemination, and size. Initially, we targeted the tweets which pass through these filters:

\begin{itemize}[noitemsep,nolistsep]
 \item Tweets in the English language.
 \item Each has an embedded URI pointing to an external resource.
 \item The embedded URI has been shortened using Bitly (bit.ly).
 \item The embedded URIs point to unique resources.
\end{itemize}

We chose the tweets which have links as the scope of the study is focused on detecting intention in sharing resources in social media. Also the shared resource provides extended 
context of the tweet making the social post more comprehensible. The reason behind choosing bitly shortened URIs is that their API provides invaluable information about the 
clicklog patterns, creation dates, rates of dissemination, and other information as will be described in the next section. Also bitly was fairly 
popular on Twitter at the time of the dataset collection (2009). To ensure our ability to collect information related to the embedded resource, we applied an extra filter ensuring that the linked resource is 
currently available on the live web (HTTP response 200 OK), at the time of the analysis, and that it is properly archived in the public archives with at least 10 mementos. 
Consequently, we extracted 5,937 unique instances to be utilized in the next stages.

\begin{figure} [ht]
\centering
\includegraphics[scale=0.6]{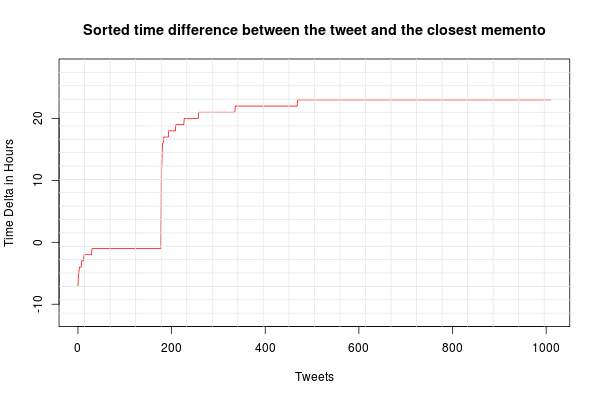}
\caption{Sorted Time delta between tweeting time and the closest memento snapshot where the negative Y axis denotes existence prior to $t_{tweet}$.}
\label{fig:tweetdelta}
\end{figure}

To create the dataset that will be processed by Mechanical Turk workers, we selected 1,124 instances randomly from the previous dataset. This training dataset will 
be assigned to the workers in the same manner to the gold standard experiment.
To have an insight of what the author was experiencing and reading upon the time of tweeting, we extracted the closest snapshot of the resource, 
to the time of the tweet, using the Memento framework. For each URI, the closest memento recorded ranged from 3.07 minutes to 56.04 hours 
from the time of the tweet, averaging 25.79 hours. Figure \ref{fig:tweetdelta} shows the difference in hours between $t_{tweet}$ and the closest memento in the public 
archives denoted by $R_{closestMemento}$. For the sake of simplicity we will consider the following approximation:
\begin{equation}
R_{closestMemento} \approx R_{tweet}
 \label{eq:closest}
\end{equation}
This shows that on average we can extract a snapshot of the state of the resource within a 
day from when the author saw it and tweeted about it. This time delta is in fact relative to the nature of the resource. In the case of continuously 
changing webpages such as CNN.com, one day will not capture everything. However, on the average, web pages are not expected to change as much within this time period. 

Along with the downloaded closest memento snapshot \\$R_{closestMemento}$, we downloaded a snapshot of the current state of the resource $R_{current}$. For the sake of simplicity as well, we consider another approximation:
\begin{equation}
R_{current} \approx R_{click}
 \label{eq:current}
\end{equation}
The agreement between Mechanical Turk workers in assigning relevancy to our training dataset of 1,124 tweets is shown in table \ref{tab:train}.
\begin{table}
\begin{center}
\begin{tabular}{|l|l|l|}
\hline
5 Turkers Agreeing (5-0 cuts) & 589 & 52.40\% \\ \hline
4 Turkers Agreeing (4-1 cuts) & 309 & 27.49\% \\ \hline
3 Turkers Agreeing (3-2 close call cuts) & 226 & 20.11\% \\ \hline \hline
Relevant Assignments & 929 & 82.65\% \\ \hline
Non-Relevant Assignments & 195 & 17.35\% \\ 
\hline
\end{tabular}
\vskip 5mm
\caption{\label{tab:train}The distribution of voting outcomes from turkers for the 1,124 assignments.}
\end{center}
\end{table}
\section{Intention Modeling}
In the previous section we collected the gold standard dataset using Mechanical Turk and tested its validity against expert opinions. Consequently, we were able to collect a larger dataset of tweets 
which have been deemed Relevant or Non-Relevant by Mechanical Turk workers as well. The dataset collected and classified contains tweets which have embedded shortened URIs or bitlys linking to a shared 
web resource. Each one of the resources is currently live and adequately covered in the public web archives at the time of this study (December 2012). 

\subsection{Feature Extraction}\label{feature}
To complement the training dataset we collected in the previous section from Mechanical Turk we explore the different angles of sharing resources in social media beyond the tweet.

\subsubsection{Link Analysis}
As mentioned earlier, most of the tweets containing resources published in 2009 include a shortened URI. One of the reasons behind this use of 
shortners is due to the space constraints of a tweet (140 characters). We extracted the tweets containing URIs shortened by bitly shortner due 
to their abundance in the SNAP dataset tweet collection. Out of the 476 million tweets in the dataset, 87 million contain bitly 
shortened URIs. The bitly API provide several parameters that could be extracted as well. The total number of clicks, hourly clicklogs, creation dates, referring websites, referring countries, and other information could also be acquired. 

The location of the resource in the domain is important. Surface web pages, as the main page or index, are different in nature from the deep web ones. Relying on the general notion that pages in the deep web are less likely to 
change as often as the root page, we need to calculate the estimated depth of the resource. Within each tweet, we expanded the resource's bitly to the original long URI and analyzed for hierarchy and depth in the web by counting the number of 
backslashes in the URI which correlates with the depth fairly well. Also we compare the lengths of the shortened URl and the original one to calculate the reduction rate. Hand in hand 
with these extracted data points, we proceed to examine the dissemination trends of that resource. 

\subsubsection{Social Media Mining}
For each 
embedded resource in a tweet, we used Topsy.com's API\footnote{http://code.google.com/p/otterapi/} to extract the total number of tweets that have been recorded linking to this resource. 
We extract the number of tweets from influential users in the Twitter-sphere as well. Finally, we downloaded the other tweets posted by different users linking to the same resource. The 
API permits us to extract a maximum of 500 tweets per resource. This collection of tweets surrounding each resource can benefit us 
in many aspects: providing extended tweet-context for the resource, showing us the social media dissemination pattern by plotting the tweet timestamps against the 
timeline, and finally, to let us examine how many of those tweets still exist and how many have been deleted.

To complete the picture, Facebook was mined as well for each of the resources in the tweets to extract the total number of shares, posts, likes, and clicks.

\subsubsection{Archival Existence}
To investigate archival existence and coverage, we calculate how many total mementos, in the aggregated public archives, are available for the resource. We record as well how many archives hold at least a copy of the resource. 
As mentioned earlier, figure \ref{fig:tweetdelta} shows the distribution of the delta time between closest archived memento and the tweet creation timestamp. Negative values on the Y-axis denote existence prior to $t_{tweet}$.

\subsubsection{Sentiment Analysis}
To go beyond the tweet text, we utilized the NLTK libraries \cite{Loper:2002:NNL:1118108.1118117} for natural language text processing to extract the most prominent sentiment in the text. For each tweet we 
extracted the positive, negative and neutral sentiment probabilities. These three probabilities give us an insight on the emotional state of the author at $t_{tweet}$.

\subsubsection{Content Similarity}
Finally, to measure the difference between the different snapshots of the resource downloaded earlier, we implemented similarity analysis functions. We transformed each of the resource's $R_{tweet}$ and $R_{click}$ to textual vectors and then calculated the cosine similarity between them. Furthermore, the collected 
tweets from Topsy.com's API associated to each resource have been accumulated in one document giving it a social context. This tweet document has been compared in similarity 
as well with $R_{tweet}$ and $R_{click}$ snapshots of the resource and the percentages were recorded. It is worth mentioning that to extract those similarities we downloaded the snapshots using the 
Lynx browser\footnote{http://lynx.browser.org/}. We used the \textit{source} option which downloads the HTML. Subsequently, on the downloaded content, we used the 
boilerplate removal from HTML pages and full text extraction algorithms by Kohlschutter et al. \cite{Kohlschutter:2010:BDU:1718487.1718542}. Finally, we calculated the cosine similarity between the each of the pairs of documents.
\subsubsection{Entity Identification}
Analyzing hundreds of tweets from Twitter timeline we noticed some interesting points. Celebrities are mentioned in abundance and have the largest number of followers. In fan tweets, most celebrities are 
mentioned by their first and last name unless they are known by only one, and finally most tweets about celebrities are in reaction or as a 
description to contemporaneous events related to the celebrity. In the field of TV, cinema, performance arts, sports, and politics, millions 
of tweets are posted daily about celebrities as a huge demographic of users use twitter as a form of news feed. Given so, we wanted to analyze the 
effect of detecting celebrity-related tweets to intention and the possibility of using it as a feature. Wikipedia has published several lists of US, 
British, and Canadian actors, and singers. Also several lists of sports players and politicians in the English speaking world. We harvested 
those lists, parsed and indexed them. Finally, given an embedded resource and upon retrieving its tweet flock 
from Topsy.com's API we test for the existence of celebrity entities in the collective tweets and record celebrity-relevance feature as true.

\subsection{Modeling and Classification}
In the features extraction phase we gathered several data points denoting context, dissemination, nature, archiving coverage, change, sentiment, and others. In this phase, we investigate which 
features have higher weights indicating importance in modeling and classifying temporal intention. We also investigate the several well known classifiers and their corresponding success rates.

In the first attempts to train the classifier and analyze the confusion matrix we noticed the instances which were classified by Mechanical Turk workers as close calls (3-2 split) 
highly populated the false positive/negative cells of the confusion matrix. These instances indicate a weak classification where one vote can deem the instance relevant or non-relevant. Thus, to 
reduce the confusion, we eliminated the training instances where this uncertainty of the workers reside. From the 1,124 instances, we kept 898 where the agreement 
on relevancy was 4 to 1, or 5 total agreement as shown in table \ref{tab:noclosecalls}. Thus, the cutoff threshold in equation \ref{eq:mt1} is increased $\textit{k}>=0.8$.
\begin{table}[ht]
\begin{center}
\begin{tabular}{|l|l|l|}
\hline
Relevant Assignments & 807 & 89.87\% \\ \hline
Non Relevant Assignments & 91 & 10.13\% \\ 
\hline
\end{tabular}
\vskip 5mm
\caption{\label{tab:noclosecalls}The distribution of voting outcomes from turkers after removing close-calls.}
\end{center}
\end{table}

Utilizing the sum of all the extracted features, we ran Weka's\footnote{http://www.cs.waikato.ac.nz/ml/weka/} different classifiers against the dataset. Subsequently, we train the model and test it using 10-fold cross validation. 
Table \ref{tab:percentages} and \ref{tab:accuracy} show 
the corresponding precisions, recalls and F-measures of the Cost Sensitive classifier based on Random Forest, which outperformed the other classifiers yielding an 90.32\% success in 
classification for our trained model.
\begin{center}
\begin{table*}[ht]
\centering
\resizebox{\textwidth}{!}{
\begin{tabular}{ |l||l|l|l|l|l| }
\hline
\multicolumn{6}{ |c| }{10-Fold Cross-Validation Testing} \\
\hline
& \textbf{Mean} & \textbf{Root Mean} & \textbf{Kappa} & \textbf{Incorrectly} & \textbf{Correctly} \\
\textbf{Classifier}  & \textbf{Absolute Error} & \textbf{Squared Error} & \textbf{Statistic} & \textbf{Classified \%} & \textbf{Classified \%} \\\hline\hline
Cost Sensitive classifier & 0.15 & 0.27 & 0.39 & 9.68\% & \textbf{90.32\%} \\
based on Random Forest &&&&& \\  \hline
\end{tabular}
}
\vskip 5mm
\caption{Results of 10-fold cross-validation against the best classifier along with the Precision, Recall and F-measure per class}
\label{tab:percentages}
\end{table*}
\end{center}
\begin{center}
\begin{table*}
\centering
\begin{tabular}{ |l||l|l|l|l| }
\hline
\textbf{Classifier} & \textbf{Precision }& \textbf{Recall }& \textbf{F-measure }& \textbf{Class}\\ \hline\hline
Cost Sensitive classifier& 0.93 & 0.96 & 0.95 & Relevant \\
based on Random Forest & 0.53 & 0.37 & 0.44 & Non-Relevant \\ \hline
Weighted Average & 0.89 & 0.90 & 0.90 &  \\ \hline
\end{tabular}
\vskip 5mm
\caption{Precision, Recall and F-measure per class}
\label{tab:accuracy}
\end{table*}
\end{center}
The classifier processed 39 different features for each instance in the training dataset. The features were collected in the feature extraction phase explained earlier in section \ref{feature}. Following 
the training phase we needed to understand the effect of each feature in the process of modeling intention. This knowledge will help us in reducing the number of required features, by the model, to estimate the intention behind a 
given social post. We applied an attribute evaluator supervised algorithm based on Ranker search method to rank the attributes or features accordingly. Analyzing the ranks, table \ref{hanyhany} 
shows the strongest six features and the order of significance in ranking the features used in classifying user temporal intention along with each's information gain.

It is also worth mentioning that using the boilerplate removal algorithm along with cosine similarity gave more significance features than HTML similarity with SimHash \cite{Charikar:2002:SET:509907.509965}.
\subsection{Evaluation}
The previous section indicates that modeling user intention via TIRM and using numerical, textual, and semantic features in a classifier 
is both feasible and accurate. In this section, we test the trained model against other tweet datasets.

\subsubsection{Extended Dataset}\label{testing}
In section \ref{gold} we extracted a dataset of 5,937 instances from which we extracted our training 1,124 
instances training dataset. The remaining 4,813 instances formed a new testing dataset. For each instance in this dataset we extracted all the features analyzed in section \ref{feature}. 
Finally, this dataset was evaluated by the trained model to test the performance and usability yielding the results in table \ref{tab:results}.
\begin{center}
\begin{table}
\centering
    \begin{tabular}{|l|l|l|}
        \hline
        Rank & Feature & Gain Ratio                                                                  \\ \hline
        1    & Existence of celebrities in Tweets&      0.149                                       \\  \hline
        2    & Number of Mementos&                 0.090                         \\  \hline 
        3    & Tweet similarity with current page &   0.071                                         \\  \hline
        4    & Similarity: Current \& Past page & 0.0527                                            \\  \hline
        5    & Similarity: Tweet and Past page&    0.04401                                        \\  \hline
        6    & Original URI's depth&           0.0324                                  \\  \hline
    \end{tabular}
\vskip 5mm
    \caption{Classifier features ordered by significance resulting from Rank Search algorithm}
    \label{hanyhany}
\end{table}
\end{center}
\subsubsection{Historical Integrity of Tweet Collections}
As described in section \ref{definition}, one of the main motives of our analysis of human intention is to maintain the historical integrity of social posts collections. 
Specifically in social posts related to historic events, preserving the consistency between the tweet and the linked resource is crucial. The link between the post and the resource 
is vulnerable to two kinds of threats: the loss of content itself (either the post or the linked resource) or the mismatch between the author's intention and what the reader is 
receiving (the resource is no longer intended by the author). In our prior work, we analyzed six datasets related to six different historic events and we evaluated how many 
of these resources are missing and how many are archived \cite{TPDL2012:Losing}. In this section, we utilize our trained model in predicting the temporal intention and in turn, in estimating the amount of mismatched 
resources where the reader is probably not reading the first draft of history intended by the tweet's author. 

Due to the nature of the collections, we limit our analysis to the resources in the form of tweets. In this case, we use the tweet datasets from the 2009-2012 events related to: Michael Jackson's Death, 
H1N1 virus outbreak, Iranian Elections, President Obama's Nobel peace prize, and the Syrian uprising. Similarly to the extended testing dataset in section \ref{testing}, we extract all the necessary 
features for each instance in the dataset. We test our model with the five datasets and 
report the results in table \ref{tab:results} as well. For each dataset we test the response headers once more to assess the percentage missing and alive, which we present in the same table. It is worth 
mentioning that when we started the experiments in September of 2012, the instances of the 3124 extended dataset were extracted to return a 200 OK response, but when we re-tested their existence 4 months later we noticed a loss of 3.23\% confirming 
the results from our previous work.
\begin{center}
\begin{table*}[t]
\centering
\resizebox{\textwidth}{!}{
\begin{tabular}{ |l||r|r||r|r| }
\hline
\textbf{Dataset} & \textbf{Status 200}& \textbf{Status 404 or Other} & \textbf{Relevant} \% & \textbf{Non-Relevant} \% \\ \hline
Extended 4,813 instances  & 96.77\% & 3.23\% & 96.74\% & 3.26\% \\ \hline \hline
MJ's Death  & 57.54\% & 42.46\% & 93.24\% & 6.76\% \\ \hline
H1N1 Outbreak & 8.96\% & 91.04\% & 97.48\% & 2.52\% \\ \hline
Iran Elections & 68.21\% & 31.79\% & 94.69\% & 5.31\% \\ \hline
Obama's Nobel & 62.86\% & 37.14\% & 93.89\% & 6.11\% \\ \hline
Syrian Uprising & 80.80\% & 19.20\% & 70.26\% & 29.75\% \\ \hline
\end{tabular}
}
\vskip 5mm
\caption{Results of testing the extended dataset \& the historic datasets in classifying relevancy along with the live percentage, and percentage missing of the resources.}
\label{tab:results}
\end{table*}
\end{center}
\subsubsection{Evaluating TIRM}
After examining the relevancy of the datasets using our developed relevancy classifier, we now use our TIRM mapping scheme in transforming the results into the intention space. The classifier was trained to be 
conservative in handling the Non-Relevant categorization. Meaning, in classifying Non-Relevancy false negatives are more tolerated than false positives (i.e., the classifier only states a resource is non-relevant 
only if it was highly confident of this estimation). Another point worth mentioning is that for our training we used the resources that are currently available on the live web; and 404 resources were not included. Table \ref{tirmresults} show the 
percentages in each of the six datasets per each class of the TIRM model after mapping relevancy to the similarity threshold of 70\%. Taking the dataset of Michael Jackson's death for example, even though the resource is still accessible nearly 3\% of the 
dataset is no longer reflecting the author's intention. It is worth noting that the results in the first quadrant of table \ref{tirmresults} are over reported. Due to the sparsity of the archives, this over reporting is essential to avoid false negatives. 
As described in figure \ref{fig:tweetdelta}, the average time delta between sharing and the closest archived version is fairly large (26 hours), in some cases the resource will keep on changing then stops after a couple of hours and stay static. Tightening 
the bounds in the same figure by more frequent archiving will lead to a large improvement in our model.

\begin{table}
\begin{center}
\begin{tabular}{rc|c}
& Relevant & Not Relevant \\
\cline{2-3}
\cline{2-3}
& MJ:41\%  & MJ:3\% \\ 
& Obama:42\% & Obama:2\% \\ 
Changed & Syria:44\% & Syria:25\% \\ 
& Iran:49\% & Iran:2\% \\ 
& H1N1:6\% & H1N1:0\% \\ 
& Extended: 53\% & Extended:2\%   \\ 
\cline{2-3}
& MJ:52\%  & MJ:4\% \\ 
& Obama:51\% & Obama:5\% \\ 
Not Changed & Syria:26\% & Syria:5\% \\ 
& Iran:46\% & Iran:3\% \\ 
& H1N1:91\% & H1N1:3\% \\ 
& Extended: 43\% & Extended:2\%   \\ 
\cline{2-3}
\end{tabular}
\vskip 5mm
\caption{\label{tirmresults}TIRM Results}
\end{center}
\end{table}
\section{Conclusions}
In this work we investigate the problem of the temporal inconsistency in social media and how it is related to the author's intention. This intention proved to be non-trivial to capture and gauge. 
Our Temporal Intention Relevancy Model successfully translated the problem of user intention to a less complicated problem of relevancy. We used Mechanical Turk to collect a gold 
standard data of user temporal intention and we verified the results by comparing the Turkers' assignments to ones conducted by experts in the field and produced a near perfect agreement. After proving the 
validity of using Mechanical Turk in data gathering, we proceeded in collecting a dataset that was used in training the classifier. We extracted several numerical, textual, and semantic features and incorporated 
them in the training dataset. The trained model is then evaluated 
against an extended larger dataset and the datasets from our previous work regarding social posts from different five historical events in the period from 2009-2012. For the shared resources, we found temporal inconsistency to range from <1\% to 25\% depending on the dataset.

For our future work, we will expand the model further more by generalizing the resources and tweets utilized in the training process, and not just the currently available and well archived resources. Also, we will increase the size of the training 
dataset and investigate the effect of each of the features and the gain resulting from combining different permutations of them.

\section{Acknowledgment}
This work was supported in part by the Library of Congress and NSF IIS-1009392.

\bibliographystyle{abbrv}

\begin{thebibliography}{10}

\bibitem{1498837}
E.~Adar, J.~Teevan, S.~T. Dumais, and J.~L. Elsas.
\newblock The web changes everything: understanding the dynamics of web
  content.
\newblock In {\em WSDM '09: Proceedings of the Second ACM International
  Conference on Web Search and Data Mining}, pages 282--291, 2009.

\bibitem{antoniades2011we}
D.~Antoniades, I.~Polakis, G.~Kontaxis, E.~Athanasopoulos, S.~Ioannidis,
  E.~Markatos, and T.~Karagiannis.
\newblock we. b: The web of short urls.
\newblock In {\em Proceedings of the 20th international conference on World
  Wide Web}, pages 715--724, 2011.

\bibitem{Bar-Yossef:2004:STG:988672.988716}
Z.~Bar-Yossef, A.~Z. Broder, R.~Kumar, and A.~Tomkins.
\newblock Sic transit gloria telae: towards an understanding of the web's
  decay.
\newblock In {\em Proceedings of the 13th international conference on World
  Wide Web}, WWW '04, pages 328--337, New York, NY, USA, 2004. ACM.

\bibitem{AWUPCP:2011:MBS}
M.~Ben~Saad and S.~Gan\c{c}arski.
\newblock Archiving the {W}eb using {P}age {C}hanges {P}attern: {A} {C}ase
  {S}tudy.
\newblock In {\em JCDL '11: Proceedings of ACM/IEEE Joint Conference on Digital
  Libraries}, Ottawa, Canada, 2011.

\bibitem{Bermingham_onusing}
A.~Bermingham and A.~F. Smeaton.
\newblock On using twitter to monitor political sentiment and predict election
  results.

\bibitem{DBLP:journals/corr/abs-1010-3003}
J.~Bollen, H.~Mao, and X.-J. Zeng.
\newblock Twitter mood predicts the stock market.
\newblock abs/1010.3003, 2010.

\bibitem{Charikar:2002:SET:509907.509965}
M.~S. Charikar.
\newblock Similarity estimation techniques from rounding algorithms.
\newblock In {\em Proceedings of the thiry-fourth annual ACM symposium on
  Theory of computing}, STOC '02, pages 380--388, New York, NY, USA, 2002. ACM.

\bibitem{Chen:2002:UIM:598693.598776}
Z.~Chen, F.~Lin, H.~Liu, Y.~Liu, W.-Y. Ma, and L.~Wenyin.
\newblock User intention modeling in web applications using data mining.
\newblock {\em World Wide Web}, 5(3):181--191, Nov. 2002.

\bibitem{857170}
J.~Cho and H.~Garcia-Molina.
\newblock Estimating frequency of change.
\newblock {\em ACM Transactions on Internet Technology}, 3(3):256--290, 2003.

\bibitem{Dai:2009:VLW:1645953.1646220}
N.~Dai and B.~D. Davison.
\newblock Vetting the links of the web.
\newblock In {\em Proceedings of the 18th ACM conference on Information and
  knowledge management}, CIKM '09, pages 1745--1748, New York, NY, USA, 2009.
  ACM.

\bibitem{996387}
Z.~Dalal, S.~Dash, P.~Dave, L.~Francisco-Revilla, R.~Furuta, U.~Karadkar, and
  F.~Shipman.
\newblock Managing distributed collections: evaluating web page changes,
  movement, and replacement.
\newblock In {\em JCDL '04: Proceedings of the 4th ACM/IEEE-CS Joint Conference
  on Digital Libraries}, pages 160--168, 2004.

\bibitem{Elsas:2010:LTD:1718487.1718489}
J.~L. Elsas and S.~T. Dumais.
\newblock Leveraging temporal dynamics of document content in relevance
  ranking.
\newblock In {\em Proceedings of the third ACM international conference on Web
  search and data mining}, WSDM '10, pages 1--10, New York, NY, USA, 2010. ACM.

\bibitem{facebook}
Facebook.com.
\newblock {Facebook official fact sheet}.
\newblock \url{http://newsroom.fb.com/content/default.aspx?NewsAreaId=22},
  2012.
\newblock [Online; accessed 17-December-2012].

\bibitem{Jansen:2007:DUI:1242572.1242739}
B.~J. Jansen, D.~L. Booth, and A.~Spink.
\newblock Determining the user intent of web search engine queries.
\newblock In {\em Proceedings of the 16th international conference on World
  Wide Web}, WWW '07, pages 1149--1150, New York, NY, USA, 2007. ACM.

\bibitem{Java:2007:WWT:1348549.1348556}
A.~Java, X.~Song, T.~Finin, and B.~Tseng.
\newblock Why we twitter: understanding microblogging usage and communities.
\newblock In {\em Proceedings of the 9th WebKDD and 1st SNA-KDD 2007 workshop
  on Web mining and social network analysis}, WebKDD/SNA-KDD '07, pages 56--65,
  New York, NY, USA, 2007. ACM.

\bibitem{Jethava:2011:SMU:2009916.2009971}
V.~Jethava, L.~Calder\'{o}n-Benavides, R.~Baeza-Yates, C.~Bhattacharyya, and
  D.~Dubhashi.
\newblock Scalable multi-dimensional user intent identification using tree
  structured distributions.
\newblock In {\em Proceedings of the 34th international ACM SIGIR conference on
  Research and development in Information Retrieval}, SIGIR '11, pages
  395--404, New York, NY, USA, 2011. ACM.

\bibitem{kahle:preserving}
B.~Kahle.
\newblock Preserving the {Internet}.
\newblock {\em Scientific American}, 276(3):82--83, March 1997.

\bibitem{quteprints47855}
A.~Kathuria, B.~J. Jansen, C.~Hafernik, and A.~Spink.
\newblock Classifying the user intent of web queries using k-means clustering.
\newblock {\em Internet Research}, 20(5):563--581, 2010.

\bibitem{klein:thesis}
M.~Klein.
\newblock {\em Using the Web Infrastructure for Real Time Recovery of Missing
  Web Pages}.
\newblock PhD thesis, Old Dominion University Department of Computer Science,
  2011.

\bibitem{Klein:2008:RLS:1429852.1429903}
M.~Klein and M.~L. Nelson.
\newblock Revisiting lexical signatures to (re-)discover web pages.
\newblock In {\em Proceedings of the 12th European conference on Research and
  Advanced Technology for Digital Libraries}, ECDL '08, pages 371--382, Berlin,
  Heidelberg, 2008. Springer-Verlag.

\bibitem{DBLP:conf/ercimdl/KleinN11}
M.~Klein and M.~L. Nelson.
\newblock Find, new, copy, web, page - tagging for the (re-)discovery of web
  pages.
\newblock In {\em Proceedings of TPDL}, pages 27--39, 2011.

\bibitem{ht10:goodtitle}
M.~Klein, J.~L. Shipman, and M.~L. Nelson.
\newblock {Is This a Good Title?}
\newblock In {\em HT '10: Proceedings of the 21st ACM Conference on Hypertext
  and Hypermedia}, pages 3--12, 2010.

\bibitem{Klein:2011:RMW:1998076.1998101}
M.~Klein, J.~Ware, and M.~L. Nelson.
\newblock Rediscovering missing web pages using link neighborhood lexical
  signatures.
\newblock In {\em Proceedings of the 11th annual international ACM/IEEE joint
  conference on Digital libraries}, JCDL '11, pages 137--140, New York, NY,
  USA, 2011. ACM.

\bibitem{Kohlschutter:2010:BDU:1718487.1718542}
C.~Kohlsch\"{u}tter, P.~Fankhauser, and W.~Nejdl.
\newblock Boilerplate detection using shallow text features.
\newblock In {\em Proceedings of the third ACM international conference on Web
  search and data mining}, WSDM '10, pages 441--450, New York, NY, USA, 2010.
  ACM.

\bibitem{Lee:2012:GGT:2232817.2232842}
J.~H. Lee and X.~Hu.
\newblock Generating ground truth for music mood classification using
  mechanical turk.
\newblock In {\em Proceedings of the 12th ACM/IEEE-CS joint conference on
  Digital Libraries}, JCDL '12, pages 129--138, New York, NY, USA, 2012. ACM.

\bibitem{Li:2008:LQI:1390334.1390393}
X.~Li, Y.-Y. Wang, and A.~Acero.
\newblock Learning query intent from regularized click graphs.
\newblock In {\em Proceedings of the 31st annual international ACM SIGIR
  conference on Research and development in information retrieval}, SIGIR '08,
  pages 339--346, New York, NY, USA, 2008. ACM.

\bibitem{Loper:2002:NNL:1118108.1118117}
E.~Loper and S.~Bird.
\newblock Nltk: the natural language toolkit.
\newblock In {\em Proceedings of the ACL-02 Workshop on Effective tools and
  methodologies for teaching natural language processing and computational
  linguistics - Volume 1}, ETMTNLP '02, pages 63--70, Stroudsburg, PA, USA,
  2002. Association for Computational Linguistics.

\bibitem{Mogadala:2012:TUB:2390131.2390145}
A.~Mogadala and V.~Varma.
\newblock Twitter user behavior understanding with mood transition prediction.
\newblock In {\em Proceedings of the 2012 workshop on Data-driven user
  behavioral modelling and mining from social media}, DUBMMSM '12, pages
  31--34, New York, NY, USA, 2012. ACM.

\bibitem{DBLP:journals/dlib/NelsonA02}
M.~L. Nelson and B.~D. Allen.
\newblock Object persistence and availability in digital libraries.
\newblock {\em D-Lib Magazine}, 8(1), 2002.

\bibitem{citeulike:336118}
M.~E.~J. Newman and J.~Park.
\newblock {Why social networks are different from other types of networks}.
\newblock {\em Physical Review E}, 68(3):036122+, sep 2003.

\bibitem{988674}
A.~Ntoulas, J.~Cho, and C.~Olston.
\newblock What's new on the web?: the evolution of the web from a search engine
  perspective.
\newblock In {\em WWW '04: Proceedings of the 13th international Conference on
  World Wide Web}, pages 1--12, 2004.

\bibitem{ws-dl:blog:losing}
H.~M. SalahEldeen.
\newblock Losing my revolution: A year after the egyptian revolution, 10\% of
  the social media documentation is gone.
\newblock
  http://ws-dl.blogspot.com/2012/02/2012-02-11-losing-my-revolution-year.html,
  2012.

\bibitem{TPDL2012:Losing}
H.~M. SalahEldeen and M.~L. Nelson.
\newblock Losing my revolution: How much social media content has been lost?
\newblock In {\em Proceedings of TPDL}, pages 125--137, 2012.

\bibitem{or2011:sanderson}
R.~Sanderson, M.~Phillips, and H.~{Van de Sompel}.
\newblock Analyzing the persistence of referenced web resources with {Memento}.
\newblock In {\em Proceedings of Open Repositories 2011}, 2011.

\bibitem{Santos:2011:ISR:2009916.2009997}
R.~L. Santos, C.~Macdonald, and I.~Ounis.
\newblock Intent-aware search result diversification.
\newblock In {\em Proceedings of the 34th international ACM SIGIR conference on
  Research and development in Information Retrieval}, SIGIR '11, pages
  595--604, New York, NY, USA, 2011. ACM.

\bibitem{Tian:2012:LCP:2339530.2339571}
Y.~Tian and J.~Zhu.
\newblock Learning from crowds in the presence of schools of thought.
\newblock In {\em Proceedings of the 18th ACM SIGKDD international conference
  on Knowledge discovery and data mining}, KDD '12, pages 226--234, New York,
  NY, USA, 2012. ACM.

\bibitem{twitter}
Twitter.com.
\newblock {Twitter numbers}.
\newblock \url{http://blog.Twitter.com/2011/03/numbers.html}, 2012.
\newblock [Online; accessed 17-December-2012].

\bibitem{nelson:memento:tr}
H.~{Van de Sompel}, M.~L. Nelson, R.~Sanderson, L.~L. Balakireva, S.~Ainsworth,
  and H.~Shankar.
\newblock {Memento: Time Travel for the Web}.
\newblock Technical Report arXiv:0911.1112, 2009.

\bibitem{Wu:2006:WWP:1143120.1143133}
M.~Wu, R.~C. Miller, and G.~Little.
\newblock Web wallet: preventing phishing attacks by revealing user intentions.
\newblock In {\em Proceedings of the second symposium on Usable privacy and
  security}, SOUPS '06, pages 102--113, New York, NY, USA, 2006. ACM.

\end{thebibliography}

\end{document}